\documentclass{article}
\usepackage{graphicx}
\usepackage{authblk}
\usepackage[margin=1in]{geometry}
\usepackage{natbib}
\usepackage{amsmath}
\usepackage{bm}
\usepackage{xcolor}
\usepackage{titlesec}
\usepackage{array}
\usepackage{tabularx}
\usepackage{booktabs}
\usepackage{siunitx}
\usepackage{tikz}
\usepackage{pgfplots}
\usepackage{tikz-3dplot}
\usepackage[hidelinks]{hyperref}
\usetikzlibrary{positioning,calc,arrows.meta}

\pgfplotsset{compat=1.18}

\tdplotsetmaincoords{60}{30}
\tdplotsetrotatedcoords{0}{0}{0}

\newcolumntype{Y}{>{\raggedright\arraybackslash}X}

\definecolor{blue_t}{RGB}{41,128,185}

\titleformat{\section}
  {\normalfont\large\bfseries\sffamily}
  {\thesection}
  {0.6em}
  {}

\titleformat{\subsection}
  {\normalfont\bfseries\sffamily}
  {\thesubsection}
  {0.6em}
  {}

\newcommand{\dd}{{\normalfont\mathrm{d}}}

\DeclareMathOperator*{\argmin}{arg\,min}

\newcommand{\uH}{\mathbf u_H}
\newcommand{\vH}{\mathbf v_H}
\newcommand{\UH}{\mathbf U_H}
\newcommand{\IH}{\mathbf I_H}

\newcommand{\JH}{\mathcal J_H}
\newcommand{\MH}{\mathcal M}
\newcommand{\XH}{\mathcal X_H}
\newcommand{\bU}{\mathbf U}
\newcommand{\bI}{\mathbf I}
\newcommand{\bZ}{\mathbf Z}
\newcommand{\Opt}{\operatorname{Opt}}
\newcommand{\refe}{\mathrm{ref}}
\newcommand{\myr}{\,\mathrm{m\,yr^{-1}}}
\newcommand{\Gs}{G^{\mathrm{s}}}
\newcommand{\Gtr}{G^{\mathrm{t}}}
\newcommand{\Go}{G^{\mathrm{o}}}
 
\title{\bfseries\sffamily\Large Physics-enriched neural solvers for transient ice-flow simulation}

\author[1]{\normalsize Thomas~Gregov}
\author[2]{\normalsize Sebastian~Rosier}
\author[1]{\normalsize Brandon~Finley}
\author[2]{\normalsize Andreas~Vieli}
\author[1]{\normalsize Guillaume~Jouvet}

\affil[1]{\normalsize Institute of Earth Surface Dynamics, Universit\'e de Lausanne, Lausanne, Switzerland}
\affil[2]{\normalsize Department of Geography, Universit\"at Z\"urich, Z\"urich, Switzerland}

\date{\normalsize \today}

\begin{document}

\maketitle

\begin{abstract} \normalsize
Transient glacier simulations with higher-order ice flow require the repeated solution of a nonlinear problem as the geometry evolves. In the online mode of the Instructed Glacier Model, the velocity field is represented by a neural network whose weights are warm-started from the previous time step and updated with a few optimizer iterations. We show that supplying the network with inexpensive input fields derived from low-order ice-flow balances improves this online solver. Unlike residual-based physics-informed neural networks, which incorporate physics through governing-equation penalties in the loss, our approach leaves the governing energy objective unchanged, adding physical structure through the network inputs. Across three real-world glacier configurations, the enriched solver is markedly more robust to solver settings. On the two alpine cases, it also improves the tuned accuracy--runtime trade-off, reducing surface-velocity errors by factors of two to four at fixed runtime and reaching few-percent relative errors with only $10^4$--$10^5$ trainable parameters, far fewer than comparable raw-input baselines. A 300-year Aletsch simulation then completes in under one minute, and the larger Valais domain in about two minutes, on a single GPU---a budget once reserved for much simpler shallow-ice models. Gains are smaller for the fast marine-terminating glacier, where nonlocal stress coupling favors larger or spectral networks. More broadly, the results suggest that enriching a neural solver's inputs with reduced-order physics can make repeated higher-order solves much cheaper, with no training data and no offline training.
\end{abstract}

\section{Introduction}
\label{sec:introduction}

\subsection{Solving ice flow with a neural network, online}
\label{subsec:intro_online}

Predicting the evolution of glaciers and ice sheets is of direct scientific and societal importance: mountain glaciers affect water resources, hazards, and ecosystems, while the Greenland and Antarctic ice sheets remain major contributors to uncertainty in future sea-level rise~\citep{Edwards2021,Zekollari2022,Rounce2023}. Producing credible predictions typically requires large ensembles of transient simulations to quantify uncertainty in climate forcing, bed topography, and basal properties. Glaciers and ice sheets are slowly evolving, strongly nonlinear mechanical systems: at glaciological time scales inertia is negligible, and ice motion is a gravity-driven creeping flow of a shear-thinning fluid---a nonlinear Stokes problem that each simulation must solve repeatedly. The higher-order mechanical models needed for such predictions in steep terrain and fast-flowing regions are, however, often too expensive at these ensemble sizes~\citep{Larour2012,Gagliardini2013}, which has motivated GPU acceleration, data-driven emulation, and physics-informed machine learning in glaciology~\citep[e.g.,][]{Raess2020,Jouvet2021,Bolibar2023, Jouvet2023,Watkins2023} and, more broadly, learned components for accelerating the solution of partial differential equations~\citep{Hsieh2019,Um2020,Kochkov2021}.

The Instructed Glacier Model (IGM) formulates glacier evolution on raster grids using differentiable, GPU-oriented operations~\citep{IGM}. Most of its components (mass balance, thermodynamics, mass conservation) are cheap local operations; the exception is ice flow, where the higher-order Blatter--Pattyn momentum balance defines a nonlinear elliptic problem that must be solved at every time step. IGM solves this problem by exploiting its variational structure: the Blatter--Pattyn velocity minimizes an energy, and instead of minimizing this energy over all velocity degrees of freedom, IGM represents the velocity field by a neural network and minimizes the same energy over the network weights, a construction known in the machine-learning literature as the Deep Ritz method~\citep{EYu2018}. Because consecutive glacier states are close, the weights obtained at one time step are reused at the next, and only a few optimizer iterations are applied per step. Each time step thus amounts to a warm-started nonlinear solve, with these iterations playing the role of the inner iterations of a classical nonlinear solver.

We refer to this as an \emph{online} neural solver: the network is part of the solver itself, re-optimized at every time step against the physical energy of the current glacier state. No training dataset and no offline training phase are involved. This is fundamentally different from the \emph{offline} use of neural networks as emulators or surrogates, which are trained once on a catalogue of solutions and then frozen~\citep{Jouvet2021,Li2021,Lu2021}. In the offline setting, the central questions are representational capacity and generalization to unseen glaciers. In the online setting treated here, the central question is different: starting from the weights of the previous time step, can a few optimizer iterations bring the network close to the new energy minimizer? A network that could represent the solution perfectly after thousands of iterations is useless online if those iterations are not affordable. The physics enrichment of \emph{offline} training objectives is addressed in a companion study~\citep{Rosier2026} and is not considered here.

\begin{figure}
    \centering
    \input{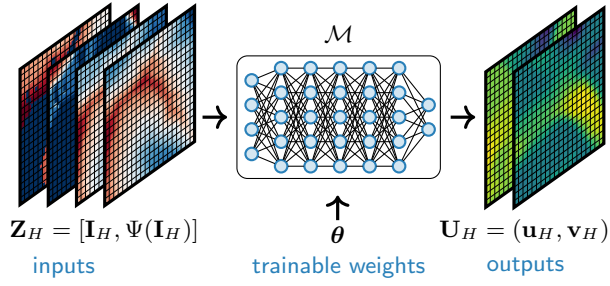}
    \caption{Schematic of the neural ice-flow parametrization used in IGM. The gridded glacier state~$\mathbf I_H=[\boldsymbol{\tau}_H,\mathbf A_H,\mathbf h_H,\mathbf s_H]$ and, for physics-enriched architectures, the derived feature fields~$\Psi(\mathbf I_H)$ (Section~\ref{sec:dahunet}) are concatenated into the network input~$\mathbf Z_H=[\mathbf I_H,\Psi(\mathbf I_H)]$; raw-input architectures use~$\mathbf Z_H=\mathbf I_H$ alone, i.e., they omit the~$\Psi$ branch. The differentiable network~$\mathcal M$, with trainable weights~$\bm\theta$, returns the horizontal velocity degrees of freedom~$\mathbf U_H=(\mathbf u_H,\mathbf v_H)$. The feature fields are deterministic functions of the state (Section~\ref{sec:dahunet}). In the online solve, the weights are updated so that the resulting velocity field decreases the discretized Blatter--Pattyn energy.}
    \label{fig:summary}
\end{figure}

\subsection{Physical structure via network inputs}
\label{subsec:intro_question}
In this online setting, the architecture is a central design choice: for a fixed energy, it determines the set of velocity fields the network can produce, the coordinates in which the minimization is carried out, and therefore how far a few optimizer iterations can go. Two architectures may represent equally accurate velocity fields after exhaustive optimization and yet behave very differently as online solvers. Section~\ref{subsec:experiments_transfer} makes this precise, separating how well an architecture's weights carry from one time step to the next from how much of the remaining error a few iterations remove.

Physical structure classically enters scientific machine learning in two ways---through the \emph{loss}, as residual penalties in physics-informed neural networks or as an energy functional in variational formulations~\citep{EYu2018,Sirignano2018,Raissi2019,Karniadakis2021,Kharazmi2021}, and through the \emph{data}, as the solution catalogues used to train operator-learning surrogates offline~\citep[e.g.,][]{Jouvet2021}. In the present solver, the loss already \emph{is} the exact physical energy and no data are involved, so both classical channels are either saturated or unavailable; in particular, the standard remedy of physics-informed learning---adding physics-based penalty terms to the loss---is not an option here without changing the problem being solved. The thesis of this paper is that a powerful third channel remains: the network's inputs. While leaving the energy, discretization, and boundary conditions strictly unchanged, one can supply the network with inexpensive fields derived from classical low-order ice-flow balances and thereby make the repeated online minimization substantially easier. It is well documented that neural networks fit some functions far more readily than others and that input transformations reshape this bias~\citep{Rahaman2019,Tancik2020}, and that physics-based losses can be hard to optimize~\citep{Krishnapriyan2021,Wang2022}. Here, we show that reshaping the network's inputs better-conditions the online minimization, so that the few per-solve iterations go much further.

We call the resulting parametrization DahuNet (\emph{DahuNet Augments Hidden Units for ice-sheet Networks}). Its added input fields are deterministic functions of the current glacier state and cost a few tensor operations per time step; the velocity used by the glacier model remains the minimizer of the full Blatter--Pattyn energy, and only the network input changes.

\subsection{Contributions and outline}
\label{subsec:intro_contributions}

We compare DahuNet against raw-input multilayer perceptron (MLP), convolutional neural network (CNN), and Fourier neural operator (FNO) baselines on three real-world glacier configurations spanning very different regimes: the Great Aletsch Glacier (slow alpine), a large multi-glacier Valais domain (many glaciers, complex margins), and Sermeq Kujalleq in Greenland (fast, marine-terminating, with a calving front). 

IGM already represents ice flow by an online, warm-started energy minimization~\citep{Jouvet2023,IGM}; we take that solver as given and ask how its network should be built. Against this background, the paper makes four contributions. (i)~We identify the network input and architecture as the primary design variable of the online solver, and introduce physics-enriched inputs (DahuNet): inexpensive fields derived from low-order ice-flow balances, with fixed, non-learned normalization, leaving the energy, discretization, and boundary conditions unchanged (Sections~\ref{sec:solver}--\ref{sec:dahunet}). (ii)~We quantify the benefit in two practical regimes---broad hyperparameter sampling (routine use) and tuned accuracy--runtime trade-offs (expert use)---across three glaciers, with surface-velocity errors reduced by factors of two to four on the alpine cases (Section~\ref{subsec:results_sweeps}). (iii)~We show that on the two alpine glaciers, physics-enriched networks solve the higher-order problem accurately with only~$10^4$--$10^5$ trainable parameters, far fewer than raw-input baselines; on the marine-terminating glacier a small network suffices only with additional optimization budget at the first stored state, and otherwise a larger or spectral network is required, which Appendix~\ref{app:locality} attributes to the nonlocal character of the flow (Sections~\ref{subsec:results_small} and~\ref{subsec:discussion_paradigm}). (iv)~We introduce a \emph{transfer test} that explains \emph{why} an architecture performs well online, by separating \emph{weight transfer} (how well the weights of one time step carry to the next) from \emph{online correction} (how much of the remaining error a few iterations remove) (Sections~\ref{subsec:experiments_transfer} and~\ref{subsec:results_mechanisms}).

The resulting picture is clear but regime-dependent: the physics-enriched family is the most reliable across all three glaciers, with the largest gains on the alpine cases; on the marine-terminating case the gains are smaller and the useful feature content must match the dominant sliding and boundary physics (Section~\ref{subsec:results_features}). Although the experiments are glaciological, the design principle is general: whenever a high-fidelity nonlinear problem must be solved repeatedly along a varying trajectory, inexpensive limiting balances can be fed to the network as inputs---rather than used as surrogates or penalty terms---to make the high-fidelity solve easier (Section~\ref{sec:discussion}).

\section{An online neural solver for higher-order ice flow}
\label{sec:solver}

This section summarizes the ice-flow formulation of IGM and defines the online neural solver. The full model is described in the model development paper~\citep{IGM}; here we only isolate the minimization problem that is common to all networks compared in this study. 

\subsection{Ice flow as energy minimization}
\label{subsec:solver_energy}

Let~$\Omega$ denote the ice domain, with surface elevation~$s$, bed elevation~$b$, and ice thickness~$h=s-b$, and let~$\bm v=(u,v,w)$ be the ice velocity. On glaciological time scales, ice behaves as an incompressible, shear-thinning viscous fluid: inertia is negligible, and Glen's flow law~\citep{Glen1955} gives an effective viscosity that decreases with the strain rate,
\begin{equation}
  \eta(\bm v)
  =
  \dfrac{1}{2}\,
  A^{-1/n}
  \left|\bm D(\bm v)\right|^{1/n-1},
  \label{eq:glen_viscosity}
\end{equation}
where~$\bm D$ is the strain-rate tensor, $A$ the Arrhenius (ice-softness) factor, and~$n>1$ Glen's exponent. This power-law rheology makes the momentum balance a nonlinear elliptic problem. IGM uses the Blatter--Pattyn approximation of the Stokes equations~\citep{Herterich1987,Blatter1995,Pattyn2003}, which retains vertical shearing and horizontal (membrane) stress gradients while assuming a cryostatic pressure; its unknowns are then the horizontal velocity components $\bm{u} = (u,v)$.

The key structural property exploited in this paper is that the Blatter--Pattyn equations are the optimality conditions of an energy~\citep{Colinge1999,Jouvet2016}: the velocity minimizes
\begin{subequations} \label{eq:continuous_energy}
\begin{align}
  \mathcal J(\bm u)
  =
  &
  \int_{\Omega}
  \frac{2A^{-1/n}}{1+1/n}
  \left|\bm D(\bm u)\right|^{1+1/n}
  \dd\Omega
  +
  \int_{\Gamma_{\mathrm b}}
  \frac{\tau_{\mathrm b}u_{\mathrm{ref}}}{1+p}
  \left(
  \frac{|\bm u|_{\mathrm b}}{u_{\mathrm{ref}}}
  \right)^{1+p}
  \dd\Gamma
  +
  \int_{\Omega}
  \rho g\nabla s\cdot\bm u
  \,\dd\Omega
  \\
  &-
  \int_{\Gamma_{\mathrm{cf}}}
  \left[
  \rho g(s-z)-p_{\mathrm w}
  \right]
  \bm u\cdot\bm n
  \,\dd\Gamma.
\end{align}
\end{subequations}
The four terms are, in order: the internal viscous dissipation; a power-law Weertman basal friction~\citep{Weertman1957}, with coefficient~$\tau_{\mathrm b}$ and exponent~$p$; the gravitational driving work; and the water-pressure contribution at the calving front~$\Gamma_{\mathrm{cf}}$, present only for marine-terminating glaciers.

IGM discretizes this problem on a regular horizontal raster grid of resolution~$H$. In this paragraph, the subscript~$H$ makes explicit that a quantity is spatially discretized on this grid. The horizontal velocity degrees of freedom are collected in~$\UH=(\uH,\vH)\in\XH$, where~$\XH$ denotes the discrete velocity space, and the gridded input fields in
\begin{equation} \label{eq:igm_inputs}
  \IH
  =
  \left[
  \boldsymbol{\tau}_H,
  \mathbf A_H,
  \mathbf h_H,
  \mathbf s_H
  \right],
\end{equation}
where~$\boldsymbol{\tau}_H$ is the basal-friction coefficient, $\mathbf A_H$ the Arrhenius factor, $\mathbf h_H$ the ice thickness, and~$\mathbf s_H$ the surface elevation. The ice-flow problem at a given glacier state is then the finite-dimensional minimization
\begin{equation} \label{eq:direct_discrete_solve}
  \UH^{\star}
  =
  \argmin_{\UH\in\XH}
  \JH(\UH;\IH),
\end{equation}
where~$\JH$ is the discretized energy. From here on, every quantity is discrete in this sense; to lighten the notation, we therefore drop the subscript~$H$ in the remainder of the paper and simply write~$\bU$, $\bI$, $\mathcal J$, $\mathcal X$, and so on, keeping in mind that all fields live on the raster grid. Problem~\eqref{eq:direct_discrete_solve} is the reference problem: it is the physical objective that all networks in this paper attempt to minimize, and solving it directly over all velocity degrees of freedom (which we do for diagnostic purposes) provides the reference solution against which they are measured.

\subsection{The neural solver}
\label{subsec:solver_neural}

Instead of optimizing over all velocity degrees of freedom, IGM represents the velocity by a neural network~$\MH$ with weights~$\bm\theta$~\citep{Jouvet2023,Cordonnier2023}, applied to an input tensor~$\bZ$:
\begin{subequations} \label{eq:neural_parametrization}
\begin{align}
  &\bU = \MH(\bm\theta;\bZ),
  \label{eq:neural_map}
  \\
  &\text{with} \quad \bZ = \bI
  \quad\text{or}\quad
  \bZ = \left[\bI,\Psi(\bI)\right].
  \label{eq:neural_inputs}
\end{align}
\end{subequations}
The two choices in~\eqref{eq:neural_inputs} distinguish the architectures compared in this paper: raw-input architectures receive the glacier state itself, while physics-enriched architectures receive the same state augmented with the physics-derived feature fields of the map~$\Psi$ defined in Section~\ref{sec:dahunet}; $[\cdot,\cdot]$ denotes channel concatenation. In both cases, the same energy is minimized over the weights,
\begin{equation}
  \min_{\bm\theta}\;
  \mathcal J\!\left(\MH(\bm\theta;\bZ);\bI\right).
  \label{eq:neural_solve}
\end{equation}
The energy, discretization, and boundary conditions are identical for all architectures; only the map from inputs to velocity differs, i.e., the network~$\MH$ and its input~$\bZ$. If~$\MH$ is the identity acting on~$\bm\theta=\bU$, problem~\eqref{eq:neural_solve} reduces to the direct solve~\eqref{eq:direct_discrete_solve}; a neural network replaces this direct search by an optimization in weight space, which can be cheaper but changes the geometry of the problem. This pipeline is the one summarized in Fig.~\ref{fig:summary}.

In a transient simulation, the rest of the model advances the glacier state, mainly through the update of the ice thickness by mass conservation, while this ice-flow minimization is repeated at every time step against the current state~$\bI_k$. Under the explicit, CFL-limited time stepping used here, each step changes the state only slightly, so consecutive states~$\bI_{k-1}$ and~$\bI_k$ are close; the weights are therefore carried over from one step to the next, and only a small, fixed number~$m$ of optimizer iterations is applied:
\begin{equation}
  \bm\theta_{k}^{(0)}
  =
  \bm\theta_{k-1}^{(m)},
  \qquad
  \bm\theta_{k}^{(m)}
  =
  \Opt_{k}^{m}(\bm\theta_{k}^{(0)}),
  \label{eq:online_update}
\end{equation}
where~$\Opt_k^m$ denotes~$m$ iterations of the optimizer applied to the energy of state~$k$. The velocity passed to the rest of the glacier model is~$\bU_k=\MH(\bm\theta_k^{(m)};\bZ_k)$. Equations~\eqref{eq:neural_parametrization}~and~\eqref{eq:online_update} define the online solver studied in this paper: a transient simulation advances two coupled sequences, the glacier states~$\bI_k$, updated by the rest of the model, and the network weights~$\bm\theta_k^{(m)}$, updated by~$m$ optimizer iterations on the energy of the current state. The coupling runs both ways: the weights produce the velocity that advances the state, and the new state defines the energy that updates the weights. Its success rests on two properties of the network: the weights of one step must remain a good starting point at the next (\emph{weight transfer}), and the few iterations must remove a large part of the remaining error (\emph{online correction}). Section~\ref{subsec:experiments_transfer} introduces a test that measures the two properties separately. Related neural solvers for time-dependent PDEs evolve the weights continuously in time to track the dynamics~\citep{Du2021,Bruna2024}; ours is the quasi-static counterpart, where the flow is inertia-free and each step is instead a static equilibrium solve, coupled to the previous one only through the warm start.

\section{Physics-enriched inputs}
\label{sec:dahunet}

DahuNet modifies a single design variable: the input tensor~$\bZ$ received by the network. This inverts the classical physics-informed strategy. PINNs improve a physics-agnostic network by adding physical terms \emph{behind} it, in the training loss. Here the objective is already the physical energy, so there is no room for that remedy; DahuNet instead adds physical information \emph{in front of} the network, in its inputs. The added fields are deterministic functions of the current glacier state that expose, before the minimization starts, the local flow directions and velocity scales suggested by classical low-order balances. They are computed by a few elementwise tensor operations per time step, contain no trainable parameters, and do not replace the higher-order solve.

\subsection{The feature fields}
\label{subsec:dahunet_features}

Let~$\Delta_x$, $\Delta_y$ denote centered finite differences on the grid. From the surface elevation we form the surface-slope fields
\begin{equation}
  \mathbf k_x = \Delta_x\mathbf s,
  \quad
  \mathbf k_y = \Delta_y\mathbf s,
  \quad
  \mathbf k=
  \left(
  \mathbf k_x^2+
  \mathbf k_y^2
  \right)^{1/2}.
  \label{eq:slopes}
\end{equation}
These are natural inputs because the slope~$\nabla s$ drives the gravitational term of the energy~\eqref{eq:continuous_energy} and thus strongly influences the local flow direction. Each field passed to the network is normalized by a fixed, non-learned scale (slope scale~$k_0=10^{-1}$, thickness scale~$h_0=200\,$m); normalized fields are denoted by a hat, e.g., $\widehat{\mathbf k}=\mathbf k/k_0$ and~$\widehat{\mathbf h}=\mathbf h/h_0$. The three-feature map contains the slope information only,
\begin{equation}
  \Psi_3(\bI)
  =
  \left[
  \widehat{\mathbf k}_x,
  \widehat{\mathbf k}_y,
  \widehat{\mathbf k}
  \right].
  \label{eq:psi3}
\end{equation}
The five-feature map appends two velocity-scale fields,
\begin{subequations} \label{eq:psi5}
\begin{align}
  \widehat{\mathbf u}_{\mathrm{sia}}
  &=
  \log\left(
  1+
  \widehat{\mathbf k}^{\,3}
  \widehat{\mathbf h}^{\,4}
  \right),
  \\
  \widehat{\mathbf u}_{\mathrm w}
  &=
  \log\left(
  1+
  \widehat{\mathbf k}^{\,3}
  \widehat{\mathbf h}^{\,3}
  \right),
  \\
  \Psi_5(\bI)
  &=
  \left[
  \widehat{\mathbf k}_x,
  \widehat{\mathbf k}_y,
  \widehat{\mathbf k},
  \widehat{\mathbf u}_{\mathrm{sia}},
  \widehat{\mathbf u}_{\mathrm w}
  \right],
\end{align}
\end{subequations}
with all operations applied elementwise. The field~$\widehat{\mathbf u}_{\mathrm{sia}}$ is, up to constants, the geometric part~$h^4|\nabla s|^3$ of the classical shallow-ice deformation-velocity scaling, and~$\widehat{\mathbf u}_{\mathrm w}$ is the corresponding geometric factor~$h^3|\nabla s|^3$ of a sliding (or membrane-flow) velocity scale; the logarithm compresses their wide dynamic range. The derivations behind these two scalings are collected in Appendix~\ref{app:feature_motivation}.

\subsection{Networks compared}
\label{subsec:dahunet_networks}

Table~\ref{tab:variants} lists the network configurations compared in this study, each defined by a backbone~$\MH$ and an input~$\bZ$. Three raw-input baselines receive~$\bZ=\bI$: a per-pixel MLP (shared weights at each grid point, no spatial coupling), a CNN (a pointwise input layer followed by a stack of two-dimensional convolutions with GELU activations and optional residual connections; its convolutions propagate information locally, which matches the spatial coherence of glacier fields), and an FNO (the standard two-dimensional Fourier neural operator~\citep{Li2021}, whose layers mix information globally through truncated spectral convolutions). The unqualified name \emph{DahuNet} refers to the CNN backbone with the five-feature map~$\Psi_5$; qualified variants (CNN+3f, FNO+3f, FNO+5f) combine either backbone with either feature map and are used for ablations, i.e., controlled comparisons that vary or corrupt one ingredient at a time.

Finally, a \emph{scrambled-feature control} tests whether any gain comes from the physical content of the features rather than from the mere presence of extra input channels. It uses the same CNN backbone and the same number of channels as DahuNet, but each feature field is spatially scrambled: a fixed randomization of its Fourier phases keeps the field's statistics but relocates its features, so high slopes and large velocity scales no longer sit where the ice actually steepens or flows fast. If this control performs poorly, extra channels alone cannot explain the DahuNet gains.

\begin{table}
  \centering
  \small
  \setlength{\tabcolsep}{4pt}
  \begin{tabular}{lll}
    \hline
    Name & Backbone & Network input~$\bZ$ \\
    \hline
    MLP & MLP & $\bI$ \\
    CNN & CNN & $\bI$ \\
    FNO & FNO & $\bI$ \\
    DahuNet & CNN & $[\bI,\Psi_5(\bI)]$ \\
    DahuNet (CNN+3f) & CNN & $[\bI,\Psi_3(\bI)]$ \\
    DahuNet (FNO+3f) & FNO & $[\bI,\Psi_3(\bI)]$ \\
    DahuNet (FNO+5f) & FNO & $[\bI,\Psi_5(\bI)]$ \\
    DahuNet (scrambled) & CNN & $[\bI,\mathcal S(\Psi_5(\bI))]$ \\
    \hline
  \end{tabular}
  \caption{Architecture naming. The main results compare MLP, CNN, FNO, and DahuNet; the DahuNet variants and the scrambled control are used for the ablation study.}
  \label{tab:variants}
\end{table}

\section{Experimental protocol}
\label{sec:experiments}

We evaluate the architectures with two complementary experiments: hyperparameter \emph{sweeps}, which measure end-to-end solver performance in transient simulations, and a \emph{transfer test}, which explains that performance by separating weight transfer from online correction. Implementation details (grids, forcings, search ranges, optimizer settings, hardware) are collected in Appendix~\ref{app:experiment_details}.

\subsection{Glacier test cases}
\label{subsec:experiments_glaciers}

\begin{figure*}
  \centering
  \includegraphics[width=\linewidth]{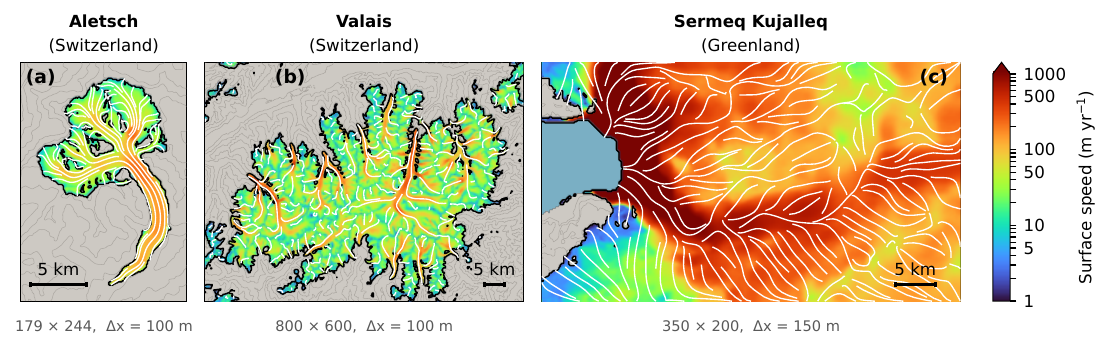}
  \caption{The three glacier test cases. Color shows the surface-speed magnitude at the maximum ice extent of the reference trajectory; white streamlines indicate the flow direction; black contours mark the glacier margins; gray isolines show the ice-free terrain. The ocean is shown in teal for Sermeq Kujalleq. Grid dimensions and horizontal resolution are indicated below each panel.}
  \label{fig:test_cases}
\end{figure*}

We use three configurations that probe different regimes (Fig.~\ref{fig:test_cases}): the Great Aletsch Glacier (Switzerland), a comparatively slow alpine valley glacier; the Valais domain (Switzerland), a large multi-glacier alpine region with faster flow; and Sermeq Kujalleq (Greenland), a fast marine-terminating outlet glacier whose energy includes the calving-front term. These cases are not a statistically exhaustive benchmark; they are chosen to test whether the same physics-enriched inputs remain useful across changes in velocity scale, margin geometry, and boundary forcing.

\subsection{Hyperparameter sweeps}
\label{subsec:experiments_sweeps}

Each sweep trial runs a full transient IGM simulation with one architecture and one draw of solver hyperparameters (network depth and width, kernel size, learning rates, number of online iterations, retraining period; see Appendix~\ref{app:experiment_sweeps}), and returns two numbers: the surface-velocity error~$E$ defined below, and the wall-clock time~$T$ per online ice-flow solve. Every trial starts from a fresh random initialization.

Two sweep types answer two practical questions. \emph{Uniform sweeps} sample the hyperparameters at random from broad admissible ranges and measure how reliably an architecture performs without tuning---the relevant property for routine, non-expert use. \emph{Pareto sweeps} use the multi-objective genetic algorithm NSGA-II~\citep{Deb2002}, through Optuna~\citep{Akiba2019}, to approximate the best achievable trade-off curve between error and runtime---the relevant property for expert use under a prescribed compute budget.

\subsection{The transfer test}
\label{subsec:experiments_transfer}

The sweeps quantify performance but not its origin. An effective online solver may succeed for two distinct reasons: because its weights, optimized at one glacier state, remain close to a minimizer at the next state (good weight transfer), or because its few online iterations reduce the energy very effectively (good online correction). The transfer test measures the two separately, on prescribed glacier states taken from a stored reference trajectory so that all architectures see exactly the same states.

For a transition between two consecutive stored states~$k\to k+1$, the test has three stages. First, the network is optimized \emph{from scratch} (fresh random weights, $2\times10^4$ iterations)on state~$k$, yielding the from-scratch energy~$J^{\mathrm s}_k$; this measures whether the architecture can solve a single state at all. Second, the optimized weights are carried unchanged to state~$k+1$ and the transferred energy~$J^{\mathrm t}_{k+1}$ is evaluated; the jump from~$J^{\mathrm s}_k$ to~$J^{\mathrm t}_{k+1}$ isolates the quality of the weight transfer. Third, a standard online budget of~$m=10$ iterations is applied at state~$k+1$, yielding the online-corrected energy~$J^{\mathrm o}_{k+1}$; the decrease from~$J^{\mathrm t}_{k+1}$ to~$J^{\mathrm o}_{k+1}$ measures the online correction.

All energies are reported as gaps relative to the reference energy~$J^{\refe}$ of the direct solve~\eqref{eq:direct_discrete_solve} at the same state. The definitions below apply to every transition~$k\to k+1$; we drop the state index to lighten the notation, writing $J^{\refe}$ for the reference energy at the state where each candidate energy is evaluated:
\begin{subequations} \label{eq:gaps}
\begin{align}
  \Gs &:= (J^{\mathrm s}-J^{\refe})/|J^{\refe}|, \\
  \Gtr &:= (J^{\mathrm t}-J^{\refe})/|J^{\refe}|, \\
  \Go &:= (J^{\mathrm o}-J^{\refe})/|J^{\refe}|.
\end{align}
\end{subequations}
The from-scratch gap~$\Gs$ is taken at state~$k$, whereas the transfer and online gaps~$\Gtr$ and~$\Go$ are taken at state~$k+1$; each gap thus measures how far a candidate energy sits above the reference at the \emph{same} state. The effectiveness of the online iterations is summarized by the \emph{reduction factor}
\begin{equation} \label{eq:recovery}
  r := \frac{\Go}{\Gtr},
\end{equation}
the factor by which a single online step contracts the transfer gap, so that smaller is better; equivalently, the step removes a fraction~$1-r$ of the gap. The factorization~$\Go=r\,\Gtr$ is the key reading aid: the final gap after an online step is exactly the product of a weight-transfer contribution ($\Gtr$) and an online-correction contribution ($r$), both small for a good solver. An architecture may therefore succeed by transferring well (small~$\Gtr$), by correcting well (small~$r$), or by both; the test tells which. Budgets and learning rates are given in Appendix~\ref{app:experiment_transfer_settings}.

\subsection{Accuracy metrics}
\label{subsec:experiments_metrics}

Velocity accuracy is measured on the horizontal surface velocity: the error~$e$ is the root-mean-square difference between the candidate and the reference surface velocities of the direct solve, taken over the glaciated area, and the sweep objective~$E$ aggregates this error over the evaluation states of the simulation. Where convenient we also quote the \emph{relative} error, i.e., $e$ divided by the root-mean-square reference surface speed over the same area.

\section{Numerical results}
\label{sec:results}

The results are organized around four findings: physics-enriched inputs make the online solver more robust and more efficient (Section~\ref{subsec:results_sweeps}); small networks are enough (Section~\ref{subsec:results_small}); the gains come from better weight transfer and, case-dependently, better online correction (Section~\ref{subsec:results_mechanisms}); and it is the physical meaning of the inputs that matters, not their number (Section~\ref{subsec:results_features}).

\subsection{Robustness and tuned accuracy--runtime performance}
\label{subsec:results_sweeps}

\begin{figure*}
    \centering
    \includegraphics[width=\textwidth]{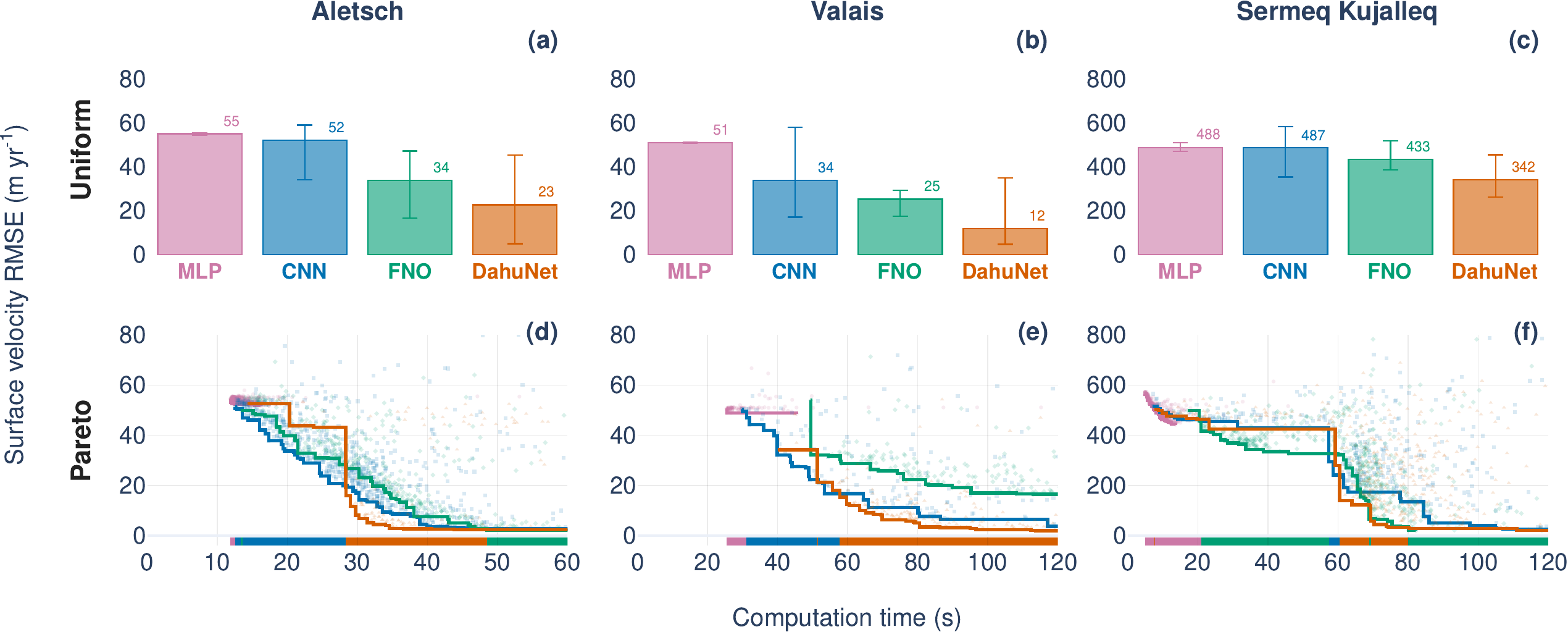}
    \caption{Solver performance in the hyperparameter sweeps, per glacier and architecture. \emph{Uniform sweeps (top row, a--c):} robustness to solver settings, shown as the median surface-velocity RMSE over the uniform sweeps, with error bars spanning the first to third quartile. \emph{Pareto sweeps (bottom row, d--f):} tuned accuracy--runtime performance, shown as surface-velocity RMSE against wall-clock time per online ice-flow solve; thin points are evaluated trials and thick staircase curves are the best achievable trade-offs (Pareto fronts) for each architecture family. DahuNet denotes the CNN backbone with the five-feature map.}
    \label{fig:sweeps}
\end{figure*}

The uniform sweeps address routine use: how well does each architecture perform when its solver hyperparameters are drawn broadly rather than tuned? Here DahuNet attains the lowest median error on all three glaciers: about~$23$, $12$, and~$342\myr$ on Aletsch, Valais, and Sermeq Kujalleq---less than half the raw-input CNN error on the alpine glaciers, and~$20$--$50\%$ below the best raw-input baseline on all three (all values in Fig.~\ref{fig:sweeps}a--c). Note that the absolute errors are not comparable across cases: Sermeq Kujalleq flows more than an order of magnitude faster than the two alpine glaciers, with surface speeds exceeding~$10^3\myr$ along its trunk (Fig.~\ref{fig:test_cases}c). Beyond improving a tuned configuration, the physics-enriched inputs make an untuned solver far less likely to fail---the property that matters for routine, non-expert use.

The Pareto sweeps address the complementary expert question: what accuracy can be reached at a prescribed runtime once the hyperparameters are tuned? On the two alpine glaciers, DahuNet occupies the low-error part of the front over the practical runtime range (Fig.~\ref{fig:sweeps}d--f). On Aletsch, it reaches a relative error of~$4\%$ ($2.3\myr$) in about $51$\,s of online solve time, where the raw-input CNN is still more than twice as inaccurate; on Valais, it reaches~$6\%$ at a runtime for which no raw-input baseline attains a comparable error, the best raw-input CNN configuration requiring roughly four times longer.

Sermeq Kujalleq also benefits from physics enrichment, but less markedly. DahuNet has the lowest median error in the uniform sweeps, yet on the tuned front the architecture families are much closer, and FNO-based variants---raw or enriched---are competitive. This is consistent with the dynamics of a fast marine-terminating outlet glacier, where longitudinal and lateral membrane stresses and the calving-front forcing make part of the solve nonlocal, so that the global spectral mixing of the FNO becomes an asset. The marine case therefore does not show unconditional dominance of a single architecture; it shows that the useful inputs and the network structure must match the governing regime, a point quantified by the ablations of Section~\ref{subsec:results_features}.

\subsection{Parameter efficiency}
\label{subsec:results_small}

\begin{figure*}
    \centering
    \includegraphics[width=\textwidth]{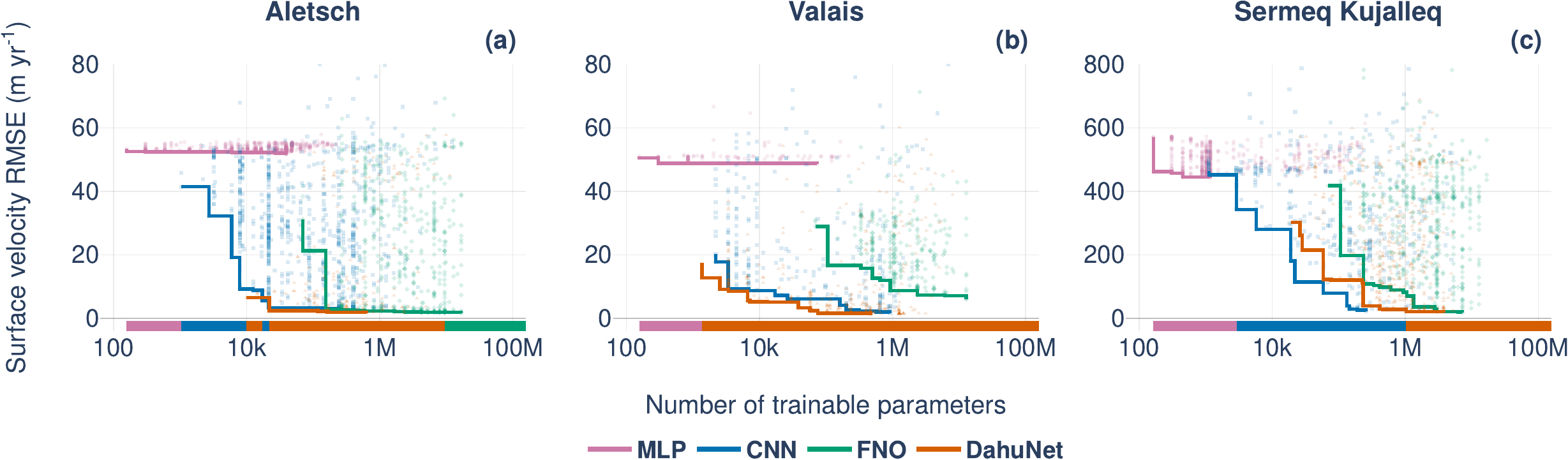}
    \caption{Accuracy versus network size. Thick staircase curves show the best surface-velocity RMSE attainable with at most a given number of trainable parameters. On the two alpine glaciers (a,~b), DahuNet reaches the low-error regime with~$10^4$--$10^5$ parameters, while the raw-input CNN and FNO improve substantially only at much larger sizes and the MLP front is nearly flat, indicating a structural rather than a capacity limitation. On the marine-terminating Sermeq Kujalleq (c), this small-network advantage is absent under the no-warm-up protocol: the raw CNN is competitive with DahuNet across~$10^4$--$10^5$ parameters, and the low-error front is reached only beyond~$\sim\!10^5$, consistent with the nonlocal flow regime discussed in Section~\ref{subsec:results_small} and Appendix~\ref{app:locality}.}
    \label{fig:params}
\end{figure*}

A central and perhaps surprising outcome of the sweeps is how small an effective online solver can be (Fig.~\ref{fig:params}). On Aletsch, a DahuNet with fewer than~$5\times10^3$ parameters already reaches a relative error of about~$10\%$, and~$4\times10^4$ parameters suffice for~$4\%$; beyond this size the accuracy saturates, improving only marginally even with fifty times more parameters. Valais behaves similarly. The raw-input baselines are very different: the plain CNN needs roughly seven times more parameters to reach a comparable accuracy on Aletsch; the FNO cannot be made smaller than about~$7\times10^4$ parameters (the spectral layers set a floor) and improves substantially only beyond~$10^6$; and the MLP never drops below the zero-velocity baseline on either glacier, at any size, confirming that some spatial coupling in the network is indispensable for this problem.

Two practical consequences follow. First, at these sizes the network itself is a minor part of the cost budget:~$4\times10^4$ single-precision parameters occupy about 160~kB, and the measured solve time grows only weakly with the parameter count (on Aletsch, multiplying the parameters by eight increases the online solve time by~$40\%$, far less than the eightfold growth of the network arithmetic). The per-iteration cost is instead dominated by contributions independent of the network size---the evaluation of the discretized energy and its adjoint on the raster grid, kernel-launch overhead, and the low GPU occupancy of small tensors---so the network forward and backward passes are not the bottleneck in this regime. Second, the comparison at fixed size is the clearest demonstration of what the physical inputs buy: at a few times~$10^3$ parameters, DahuNet achieves a relative error of about~$10\%$ while the plain CNN is still above~$50\%$. Section~\ref{subsec:discussion_paradigm} discusses what this parameter efficiency implies for the cost of higher-order ice-flow modeling at large.

The marine-terminating Sermeq Kujalleq behaves differently. Under the no-warm-up protocol the small-network advantage does not hold: the raw CNN is competitive with DahuNet throughout~$10^4$--$10^5$ parameters, and a low relative error is reached only once the network grows beyond~$\sim\!10^5$ parameters (Fig.~\ref{fig:params}c). This is the parameter-axis signature of the same nonlocality seen in the tuned sweeps (Section~\ref{subsec:results_sweeps}) and the transfer test (Section~\ref{subsec:results_mechanisms}). The velocity-scale features supply the correct \emph{local} magnitude of the flow, but in the membrane-stress regime that governs a tidewater trunk the surface velocity is a \emph{nonlocal} functional of the driving stress, coupled over many ice thicknesses; a convolutional network must span that coupling length through its receptive field, which grows only with depth, so more parameters---or the globally coupled spectral layers of an FNO---are needed. The small-network regime is thus a property of glaciers whose leading-order flow is itself local, the shallow-ice-dominated alpine cases, rather than a universal consequence of the physics-enriched inputs. Appendix~\ref{app:locality} develops this argument. The transfer test already shows that DahuNet represents the marine state well from scratch (Table~\ref{tab:transfer_medians}), and a controlled pretraining experiment shows that extra optimization budget recovers the small-network regime. The obstacle is therefore budget, not capacity.

\subsection{Attribution of the gains: weight transfer and online correction}
\label{subsec:results_mechanisms}

\begin{figure*}
    \centering
    \includegraphics[width=\linewidth]{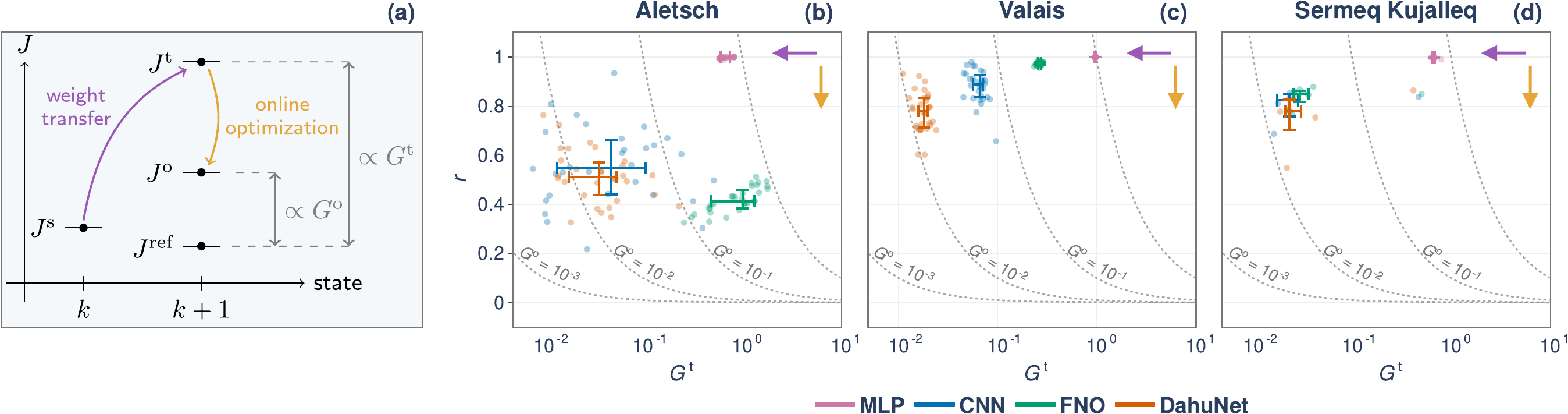}
    \caption{The two mechanisms of an effective online solve, measured by the transfer test.
    \emph{(a)} One transition $k\to k+1$: carrying the weights from state~$k$ raises the energy from~$J^{\mathrm s}$ to~$J^{\mathrm t}$ (\emph{weight transfer}, purple), and the ten online iterations lower it to~$J^{\mathrm o}$ (\emph{online optimization}, gold); gaps are taken relative to the direct-solve reference~$J^{\refe}$.
    \emph{(b--d)} Faint points are individual transitions; thick crosses are family medians with first-to-third-quartile whiskers. Smaller~$\Gtr$ means the weights of one state remain closer to a minimizer at the next; smaller~$r=\Go/\Gtr$ means the ten iterations remove more of the remaining gap. Arrows mark the improvement direction of each mechanism, colored as in~(a). Dotted curves are isolines of~$\Go=r\,\Gtr$: an architecture improves by moving toward the lower-left corner.}
    \label{fig:mechanism_plane}
\end{figure*}

The transfer test explains the sweep rankings; the complete medians are given in Table~\ref{tab:transfer_medians} of Appendix~\ref{app:extended_diagnostics}, and Fig.~\ref{fig:mechanism_plane}b--d displays each glacier in the plane spanned by the two mechanisms. The configurations analyzed are representative tuned solvers selected from the Pareto sweeps (Appendix~\ref{app:experiment_transfer_settings}).

On Aletsch, both mechanisms favor DahuNet: its weights transfer better between states and the ten iterations recover slightly more of the remaining gap, which explains its strong advantage on the Aletsch front. The raw-input FNO exposes the opposite failure mode: it solves a single state very well from scratch ($\Gs\sim10^{-3}$), but the transfer is catastrophic ($\Gtr\approx1$, a jump of almost three orders of magnitude)---the globally mixing spectral layers are not robust to the localized change of the glacier state between steps---and even recovering more than half of that gap within the ten iterations cannot compensate for so poor a starting point. Being able to represent the solution is thus necessary but not sufficient online.

On Valais, the prescribed state increment is too small to expose a transfer penalty: for all families the transfer gap essentially equals the from-scratch gap, and the reduction factors are close to one, i.e., the online step barely reduces the gap (Table~\ref{tab:transfer_medians}). Here the DahuNet advantage is of a third kind: it simply reaches a better solution from scratch, and the transfer preserves that advantage.

On Sermeq Kujalleq, all converging spatial architectures cluster in a small region of the plane and reach similar final gaps: the raw CNN and DahuNet share the same transfer gap and end at indistinguishable $\Go$, DahuNet contracting the gap slightly more per online solve (lower $r$). The surface-velocity errors of all converging architectures are nearly indistinguishable in this controlled test, indicating that energy gap and velocity error are less tightly coupled in the marine regime; the test should be read there as a mechanism attribution rather than as a ranking. Consistently with these regimes, the online corrections concentrate in the glacier interior on Aletsch and near the marine margin on Sermeq Kujalleq: the enrichment helps precisely where its coordinates match the active physics.

In summary, DahuNet's online advantage has a different origin in each regime---better weight transfer on Aletsch, a better from-scratch solution on Valais, convergence at all on the marine trunk---but the factorization~$\Go=r\,\Gtr$ localizes it in every case. Weight transfer and online correction exist only online: an offline emulator, frozen after training, lives entirely on the from-scratch axis, so representational capacity alone is a poor predictor of online performance.

\subsection{Effect of feature content versus channel count}
\label{subsec:results_features}

The scrambled-feature control (Section~\ref{subsec:dahunet_networks}; same backbone and channel count, but each feature's spatial structure destroyed by phase randomization) performs far worse than even the raw-input CNN: it fails the from-scratch solve on Aletsch and Sermeq Kujalleq, and its velocity errors are roughly an order of magnitude above DahuNet's on the alpine glaciers (Table~\ref{tab:transfer_medians}). Misleading inputs are therefore not a benign form of extra capacity: they actively degrade the optimization. In the sweeps, the scrambled control needs an order of magnitude more parameters than DahuNet to reach a comparable accuracy on Aletsch (Fig.~\ref{fig:app_variant_bars}).

The comparison between the three- and five-feature maps shows, in turn, that the useful feature content depends on the glacier regime. On Aletsch, the slope features alone essentially suffice (the 3f and 5f variants are nearly tied in the transfer test). On Valais, the five-feature map is clearly preferable, with an energy gap three times smaller and half the velocity error. On Sermeq Kujalleq the difference is qualitative: the 3f variant fails the from-scratch solve altogether, while the 5f variant converges (Table~\ref{tab:transfer_medians}). In the tidewater regime, where sliding and membrane stresses dominate, slope information alone does not contain the required velocity scale, and the two proxy fields~$\widehat{\mathbf u}_{\mathrm{sia}}$ and~$\widehat{\mathbf u}_{\mathrm w}$ supply exactly the missing information. For broad use across the tested regimes, the CNN backbone with the five-feature map is the most robust compromise; more specialized combinations (notably FNO-based variants on the marine case) can outperform it on individual glaciers.

\section{Discussion and conclusions}
\label{sec:discussion}

\subsection{Summary of findings}
\label{subsec:discussion_summary}

The experiments isolate a single design variable---what the network receives as input---and show that it alone materially changes the behavior of an online neural ice-flow solver. With five inexpensive physics-derived input fields, and without touching the energy, discretization, or boundary conditions, the solver becomes markedly more robust to hyperparameter choices on all three glaciers. After tuning it is also more efficient, with surface-velocity errors reduced by factors of two to four on the alpine cases at fixed runtime. Since every architecture minimizes the same discrete energy, these differences are computational, not physical: they do not change the solution, only how readily the online optimization reaches it---the warm start it begins from, the velocity fields a few iterations can actually reach, and the cost of each step.
The transfer test makes the attribution quantitative through the factorization~$\Go=r\,\Gtr$: on Aletsch the enriched inputs mainly improve the weight transfer between time steps; on Valais they mainly improve the solution reached from scratch, which the transfer then preserves; on the marine-terminating case they are a prerequisite for convergence at all (through the velocity-scale fields), while the choice of backbone matters more because part of the solve is nonlocal. The scrambled control shows that none of this is an effect of extra channels or parameters: features with the right magnitudes but their spatial structure scrambled perform worse than no features at all. Loosely speaking, the feature map acts as a state-dependent change of coordinates that preconditions the optimization---not as added capacity.

\subsection{Higher-order ice flow becomes affordable online}
\label{subsec:discussion_paradigm}

The parameter-efficiency result of Section~\ref{subsec:results_small} has a direct practical consequence: for transient single-glacier simulations, higher-order ice flow is affordable online. A DahuNet with a few times~$10^4$ parameters occupies a fraction of a megabyte and, updated online with about ten Adam iterations per time step, carries the entire 300-year Aletsch simulation in under a minute---and the much larger multi-glacier Valais domain in about two minutes---on a single consumer GPU, at a few percent relative error. At such absolute costs, the traditional argument for shallow-ice models, namely that solving the higher-order equations along a transient trajectory is unaffordable, largely disappears: within a computing budget for which one would classically have accepted the shallow-ice approximation, the online solver delivers the higher-order solution. This shift is enabled jointly by the (variational) online formulation, by GPU efficiency at small network sizes, and---as the ablations show---by the physical input features, without which the small-network regime is simply not reachable.

This removes much of the classical cost--accuracy trade-off between shallow and higher-order ice-flow models, and it does so in a way that differs qualitatively from offline emulation. Offline surrogates and operator-learning models typically require~$10^6$ or more parameters, large training catalogues, an expensive training phase, and carry a generalization risk when applied outside their training distribution~\citep[e.g.,][]{Jouvet2021}. The online solver studied here uses no data and no training phase, adapts to the current glacier state by construction, and achieves its accuracy with two orders of magnitude fewer parameters---because the physical input features carry the leading-order structure that a raw-input network would otherwise have to discover, and because at these sizes the per-step cost of the network on a GPU is small (Section~\ref{subsec:results_small}). Parameter count is costlier online than offline: an offline emulator pays for its size once at training and then runs forward-only, whereas the online solver performs a backward pass at every time step. Small networks are therefore especially valuable here, and the physical inputs are what let a small network solve the problem accurately (Section~\ref{subsec:results_small}).

\subsection{Guidance for building physics-enriched inputs}
\label{subsec:discussion_guidance}

Nothing in the construction is specific to glaciology beyond the availability of inexpensive limiting balances, which suggests a general design principle: a neural solver can be enriched with physics through its inputs whenever reduced-order or asymptotic approximations of the solution scales are available---in nonlinear elasticity with membrane or shell limits, porous-media flow with Darcy or lubrication scalings, or geophysical flows with shallow or hydrostatic reductions. More broadly, this argues for treating the input representation of a neural solver as a numerical-analysis object in its own right: just as the choice of basis and preconditioner determines the cost of a classical solve, the choice of network inputs determines the cost of an online neural one. Physical structure can enter a neural solver not only through what is minimized, but through the coordinates in which the minimization takes place. The experiments translate this principle into concrete guidance. The target is the smallest network that makes the minimizer reachable within the per-step iteration budget, not the largest network that could represent it. The input features should expose the expected local directions and magnitudes of the solution without prescribing it; they should be nondimensionalized with fixed, a priori scales and log-compressed when their dynamic range is wide; and divisions by fields that may vanish should be avoided even when they would make a proxy dimensionally exact. The useful feature content is regime-dependent and must include the dominant balance---on the tidewater case, slope information alone is insufficient and the velocity-scale fields are essential. Finally, a phase-scrambling control provides an inexpensive test that observed gains reflect physical meaning rather than added capacity.

\subsection{Limitations and outlook}
\label{subsec:discussion_outlook}

The study covers three glacier configurations and uses Adam as the online optimizer. Whether the same rankings persist under quasi-Newton or natural-gradient updates is open; energy natural gradient descent, in particular, can substantially improve this class of minimizations~\citep{Muller2023,Guzman2025}, but is more expensive per iteration, and under the explicit, CFL-limited time stepping of the present glacier model the ice-flow solve is called very often. A fair comparison should therefore couple the choice of optimizer with the choice of time integration: if implicit schemes allow larger time steps, costlier per-step updates could be amortized. A second direction is adaptive retraining, in which the online iteration budget~$m$ and the retraining frequency are set automatically rather than fixed in advance, spending effort only when weight transfer degrades; this needs a cheap online signal of solver health, for which the per-step energy decrease is a natural candidate. A third direction is to combine the offline and online regimes without conflating them, for example by using a pretrained offline emulator~\citep{Rosier2026} as an initialization stage that the physics-enriched online solver then specializes to the current state, possibly with residual-correction stages in the spirit of multistage networks~\citep{Wang2024}. 

\appendix

\section{Physical motivation of the feature fields}
\label{app:feature_motivation}

This appendix motivates the two velocity-scale fields of the five-feature map~\eqref{eq:psi5} from limiting balances of the ice-flow problem, and records the normalization choices.

\subsection{Shallow-ice deformation scale}
In the shallow-ice limit, the dominant shear stress at depth~$z$ scales as~$\rho g(s-z)|\nabla s|$. Combining this stress scale with Glen's law~\eqref{eq:glen_viscosity} and integrating the resulting shear rate through the ice column gives the classical depth-averaged deformation-velocity scaling~\citep{Hutter1983,Greve2009}
\begin{equation}
  |\bar{\bm u}|
  \propto
  A(\rho g)^n h^{n+1}|\nabla s|^n .
  \label{eq:app_sia_scaling}
\end{equation}
For~$n=3$ the geometric part is~$h^4|\nabla s|^3$, which motivates~$\widehat{\mathbf u}_{\mathrm{sia}}$ in~\eqref{eq:psi5}. The constants and the Arrhenius factor are deliberately not duplicated in the feature, because~$\mathbf A$ is already part of the raw state~$\bI$.

\subsection{Sliding and membrane-flow scale}
The second field represents a driving-stress scale raised to Glen's exponent, and can be motivated from two limiting balances. In a local sliding balance, the traction that basal resistance must support scales as the driving stress
\begin{equation}
  \tau_\mathrm{d}
  :=
  \rho g h |\nabla s| ,
  \label{eq:app_driving_stress}
\end{equation}
and in the velocity form of Weertman sliding the basal speed grows as a power of the ratio between this scale and the friction coefficient,~$|\bm u| \sim u_{\mathrm{ref}}(\tau_\mathrm{d}/\tau_{\mathrm b})^p$. For the cubic law used here, the geometric factor is~$h^3|\nabla s|^3$. The same factor arises in a membrane-dominated regime: over a weakly buttressed ice shelf, basal resistance is negligible and the driving stress is balanced by membrane stresses~\citep{MacAyeal1989}, so that the extensional strain rate scales as~$A\tau_\mathrm{d}^{\,n}\propto A(\rho g)^n h^n|\nabla s|^n$. A complete shelf velocity scale would additionally involve a horizontal integration length and boundary conditions, but the local state-dependent factor exposed to the network is again~$h^3|\nabla s|^3$ for~$n=3$; this motivates~$\widehat{\mathbf u}_{\mathrm w}$.

\subsection{Normalization and robustness choices}
All feature fields are normalized by fixed, non-learned scales ($k_0=10^{-1}$ for slopes, $h_0=200\,$m for thickness) so that their magnitudes are comparable to those of the raw channels, and the two velocity proxies are log-compressed because their dynamic range would otherwise span several orders of magnitude. A dimensional Weertman proxy would divide by the basal-friction field~$\boldsymbol\tau$; we deliberately avoid this because~$\boldsymbol\tau$ may vanish in parts of the domain and the division is numerically fragile. Instead, $\widehat{\mathbf u}_{\mathrm w}$ indicates where the driving stress and thickness make large sliding or membrane-dominated velocities plausible, and the friction field remains available to the network separately within~$\bI$. Neither proxy prescribes the solution: the velocity used by the model is always the minimizer of the Blatter--Pattyn energy.

\section{Experimental details}
\label{app:experiment_details}

This appendix records the settings needed to reproduce the experiments of Section~\ref{sec:experiments}. All runs were performed on NVIDIA GeForce RTX 4090 GPUs.

\subsection{Glacier configurations}
\label{app:experiment_glaciers}

Table~\ref{tab:app_glaciers} summarizes the three configurations and their reference trajectories.

\begin{table*}
  \centering
  \footnotesize
  \setlength{\tabcolsep}{5pt}
  \renewcommand{\arraystretch}{1.12}
  \begin{tabularx}{\textwidth}{@{}>{\raggedright\arraybackslash}p{0.245\textwidth}YYY@{}}
    \toprule
    \textbf{Glacier} & \textbf{Aletsch} & \textbf{Valais} & \textbf{Sermeq Kujalleq} \\
    \midrule
    Type & Land-terminated & Land-terminated & Marine-terminated \\
    Grid size & $179\times244$ & $800\times600$ & $350\times200$ \\
    Resolution~$H$ (m) & $100$ & $100$ & $150$ \\
    Forcing & Atmospheric: change in equilibrium-line altitude & Atmospheric: change in equilibrium-line altitude & Calving: change in calving-front position \\
    Simulation period (yr) & $[200,500]$ & $[200,500]$ & $[0,10]$ \\
    Transfer-test states (yr) & $\{200,210,\ldots,490\}$ & $\{200,210,\ldots,490\}$ & $\{1,2,\ldots,9\}$ \\
    Transfer-test time step~$\Delta t$ (yr) & $0.1$ & $0.05$ & $0.05$ \\
    \bottomrule
  \end{tabularx}
  \caption{Glacier-specific settings used in the sweeps and in the prescribed-state transfer test.}
  \label{tab:app_glaciers}
\end{table*}

\subsection{Sweep protocols and search space}
\label{app:experiment_sweeps}

Uniform sweeps contain 50 trials per architecture and glacier; Pareto sweeps contain up to 500 trials per architecture and glacier and use NSGA-II with population size~20 through Optuna~\citep{Deb2002,Akiba2019}. The two objectives are the surface-velocity RMSE~$E$ defined in Section~\ref{subsec:experiments_metrics} and the measured wall-clock time~$T$ per online ice-flow solve.

The common search space is given in Table~\ref{tab:app_search_space}; parameters that do not apply to a backbone are ignored (e.g., the convolutional kernel size for the MLP). The same ranges are used for the raw-input baselines, the DahuNet variants, and the scrambled control. All updates use Adam~\citep{Kingma2015} with~$(\beta_1,\beta_2)=(0.8,0.9995)$; weights are initialized with Glorot-uniform initialization; all backbones use GELU activations. The ice-flow discretization uses two vertical velocity layers, bilinear horizontal basis functions, the MOLHO vertical basis~\citep{dosSantos2022}, Glen exponent~$n=3$, and single-precision arithmetic.

\begin{table*}
  \centering
  \footnotesize
  \setlength{\tabcolsep}{5pt}
  \renewcommand{\arraystretch}{1.12}
  \begin{tabularx}{\textwidth}{@{}>{\raggedright\arraybackslash}p{0.25\textwidth}>{\ttfamily\raggedright\arraybackslash}p{0.18\textwidth}YY@{}}
    \toprule
    \textbf{Parameter} & \normalfont\textbf{Code name} & \textbf{Range or choices} & \textbf{Sampling rule} \\
    \midrule
    Initial optimizer steps & nbit\_init & $[100,3000]$ & Integer, step 100 \\
    Online optimizer steps & nbit & $[1,20]$ & Integer \\
    Full-retraining period & retrain\_freq & $[1,50]$ & Integer \\
    Online learning rate & lr & $[5\times10^{-5},5\times10^{-3}]$ & Log-uniform \\
    Initialization learning rate & lr\_init & $[10^{-4},10^{-2}]$ & Log-uniform \\
    Number of hidden layers & nb\_layers & $[2,16]$ & Integer, step 2 \\
    Hidden width & nb\_out\_filter & $[8,128]$ & Integer, step 8 \\
    CNN kernel size & conv\_ker\_size & $\{3,5,7\}$ & Categorical \\
    Residual connections & residual & $\{\text{True}, \text{False}\}$ & Categorical \\
    FNO Fourier modes & modes1 & $[4,16]$ & Integer \\
    FNO channel width & width & $[16,64]$ & Integer, step 8 \\
    \bottomrule
  \end{tabularx}
  \caption{Hyperparameter ranges used in the uniform and Pareto sweeps.}
  \label{tab:app_search_space}
\end{table*}

\subsection{Transfer-test settings}
\label{app:experiment_transfer_settings}

For each architecture and glacier, one representative configuration is selected from the Pareto sweep: the lowest-RMSE trial satisfying the online-time constraint~$T\le 30$\,s (Aletsch), $T\le 80\,$s (Valais), $T\le 60\,$s (Sermeq Kujalleq). This fixes the architecture hyperparameters before the transfer test is run.

The from-scratch solve uses~$2\times10^4$ Adam iterations, with a glacier-specific but architecture-independent learning rate ($\alpha^{\mathrm s}=2\times10^{-4}$ on Aletsch, $2\times10^{-3}$ on Valais, $1\times10^{-3}$ on Sermeq Kujalleq); using a single value per glacier makes the comparison attributable to the parametrization rather than to per-architecture optimizer tuning. The online correction uses the same budget and learning rate for all glaciers and architectures, $m=10$ and~$\alpha=5\times10^{-5}$. The reference energy~$J^{\refe}_k$ is computed independently of any network by optimizing directly over the velocity degrees of freedom: $20{,}000$ Adam iterations with learning rate~$0.9$, followed by L-BFGS with an Armijo line search for at most~$5000$ iterations and a relative stopping tolerance of~$10^{-8}$. Because this reference is itself computed to finite tolerance, gaps marginally below zero can occur when a network solution falls slightly below the numerical reference energy (see Table~\ref{tab:transfer_medians}).

\section{Complete transfer-test results and variant sweeps}
\label{app:extended_diagnostics}

Table~\ref{tab:transfer_medians} reports the transfer-test medians for all raw-input baselines, DahuNet variants, and the scrambled-feature control; the main text quotes only the values needed for the mechanism attribution. The table separates two failure modes that are otherwise easy to conflate. On Aletsch, the FNO-based models solve the from-scratch problem well but suffer a large transfer gap: their weakness is the weight transfer. On Sermeq Kujalleq, the CNN+3f and scrambled variants already fail the from-scratch solve, so their poor post-transfer values do not diagnose the transfer at all; they diagnose an inadequate parametrization for the marine regime.

\begin{table*}
  \centering
  \footnotesize
  \setlength{\tabcolsep}{14pt}
  \renewcommand{\arraystretch}{1.10}
  \begin{tabularx}{\textwidth}{@{}>{\raggedright\arraybackslash}X
    S[scientific-notation=true,table-format=-1.2e1]
    S[scientific-notation=true,table-format=1.2e1]
    S[scientific-notation=true,table-format=1.2e1]
    S[table-format=1.3]
    S[scientific-notation=true,table-format=1.2e1]@{}}
    \toprule
    \textbf{Architecture} & {$\Gs$} & {$\Gtr$} & {$\Go$} & {$r$} & {$e$~$(\mathrm{m\,yr^{-1}})$}\\
    \midrule
    \multicolumn{6}{@{}l}{\textit{Aletsch}} \\
    \midrule
    MLP & 6.32e-1 & 6.44e-1 & 6.42e-1 & 0.999 & 26.2 \\
    CNN & 6.24e-3 & 4.80e-2 & 2.86e-2 & 0.548 & 1.98 \\
    FNO & 1.72e-3 & 1.01e0 & 4.24e-1 & 0.413 & 1.98 \\
    DahuNet (CNN+5f) & -1.3e-3 & 3.62e-2 & 1.73e-2 & 0.512 & 1.30 \\
    DahuNet (CNN+3f) & -1.2e-3 & 2.75e-2 & 1.43e-2 & 0.446 & 1.39 \\
    DahuNet (FNO+3f) & 6.75e-4 & 1.12e0 & 3.27e-1 & 0.286 & 2.12 \\
    DahuNet (FNO+5f) & 3.98e-4 & 1.44e0 & 6.81e-1 & 0.444 & 2.19 \\
    DahuNet (scrambled) & 1.84e0 & 2.04e0 & 1.99e0 & 0.962 & 23.4 \\
    \addlinespace[0.35em]
    \midrule
    \multicolumn{6}{@{}l}{\textit{Valais}} \\
    \midrule
    MLP & 9.76e-1 & 9.76e-1 & 9.76e-1 & 1.000 & 51.7 \\
    CNN & 6.61e-2 & 6.66e-2 & 5.64e-2 & 0.889 & 6.78 \\
    FNO & 2.67e-1 & 2.67e-1 & 2.59e-1 & 0.974 & 14.3 \\
    DahuNet (CNN+5f) & 1.76e-2 & 1.81e-2 & 1.38e-2 & 0.780 & 2.33 \\
    DahuNet (CNN+3f) & 5.04e-2 & 5.10e-2 & 4.47e-2 & 0.899 & 4.69 \\
    DahuNet (FNO+3f) & 6.04e-2 & 6.07e-2 & 5.81e-2 & 0.956 & 4.25 \\
    DahuNet (FNO+5f) & 1.30e-1 & 1.31e-1 & 1.26e-1 & 0.973 & 8.46 \\
    DahuNet (scrambled) & 2.06e-1 & 2.06e-1 & 1.99e-1 & 0.979 & 14.6 \\
    \addlinespace[0.35em]
    \midrule
    \multicolumn{6}{@{}l}{\textit{Sermeq Kujalleq}} \\
    \midrule
    MLP & 5.99e-1 & 6.02e-1 & 6.01e-1 & 0.999 & 487 \\
    CNN & 3.15e-3 & 2.14e-2 & 1.49e-2 & 0.666 & 405 \\
    FNO & 2.98e-3 & 2.42e-2 & 1.99e-2 & 0.803 & 404 \\
    DahuNet (CNN+5f) & 2.15e-3 & 2.15e-2 & 1.59e-2 & 0.707 & 406 \\
    DahuNet (CNN+3f) & 1.10e0 & 1.16e0 & 1.15e0 & 0.997 & 484 \\
    DahuNet (FNO+3f) & 2.48e-3 & 2.34e-2 & 1.91e-2 & 0.791 & 404 \\
    DahuNet (FNO+5f) & 4.89e-3 & 2.73e-2 & 2.16e-2 & 0.766 & 405 \\
    DahuNet (scrambled) & 1.43e0 & 1.46e0 & 1.44e0 & 0.996 & 437 \\
    \bottomrule
  \end{tabularx}
  \caption{Transfer-test medians: from-scratch gap~$\Gs$, transfer gap~$\Gtr$, gap after the ten online iterations~$\Go$, reduction factor~$r$, and surface-velocity RMSE~$e$ after the online iterations. Each column is a median taken independently over transitions, so the factorization~$\Go=r\,\Gtr$ holds for every transition but not exactly on the tabulated median row.}
  \label{tab:transfer_medians}
\end{table*}

\begin{figure*}
    \centering
    \includegraphics[width=0.95\textwidth]{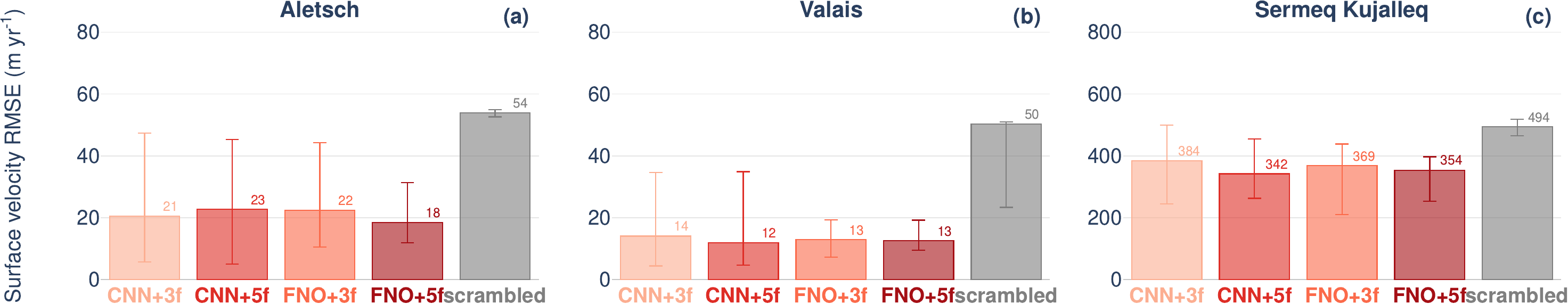}
    \caption{Robustness of the DahuNet variants and of the scrambled-feature control in the uniform sweeps: median surface-velocity RMSE per glacier and variant; error bars span the first to third quartile.}
    \label{fig:app_variant_bars}
\end{figure*}

Figure~\ref{fig:app_variant_bars} complements the main-family comparison of Fig.~\ref{fig:sweeps}a--c with the uniform-sweep robustness of the DahuNet variants and the scrambled control. Its main features are consistent with the ablation picture of Section~\ref{subsec:results_features}: all physics-enriched variants outperform the scrambled control by a wide margin on every glacier, and on Sermeq Kujalleq the CNN+5f variant attains the lowest median error.

\section{Flow locality and the small-network regime}
\label{app:locality}

The small-network result of Section~\ref{subsec:results_small} is clean on the two alpine glaciers but not on the marine-terminating Sermeq Kujalleq under the no-warm-up protocol, where the raw CNN is competitive with DahuNet across~$10^4$--$10^5$ parameters and the low-error front is reached only beyond~$\sim\!10^5$ (Fig.~\ref{fig:params}c). This appendix explains that contrast in terms of the locality of the underlying flow, and shows that the marine obstacle is optimization budget rather than representational capacity.

\subsection{Local versus nonlocal flow}
The velocity-scale features~$\widehat{\mathbf u}_{\mathrm{sia}}$ and~$\widehat{\mathbf u}_{\mathrm w}$ are pointwise-local: the value at a grid point depends only on the thickness and surface slope at that point. In the shallow-ice regime that dominates the grounded alpine glaciers, the depth-averaged velocity genuinely is, to leading order, a local function of~$(h,\nabla s)$ (Appendix~\ref{app:feature_motivation}); the feature is then almost the answer, and a small network with little training converts it into an accurate field. In the membrane-stress regime that governs a fast tidewater trunk, the surface velocity is instead a \emph{nonlocal} functional of the driving stress: longitudinal and lateral coupling redistribute stress over many ice thicknesses. The local feature then supplies approximately the right magnitude and pattern but leaves them spatially unpropagated, and the network's task shifts from reading off a local map to learning the propagation operator that spreads the local signal over the coupling length.

This single distinction organizes the parameter-axis observations. A raw CNN must span the coupling length through its receptive field, which grows only with depth~$\times$ kernel size, so at Sermeq Kujalleq it needs a network of order~$10^5$ parameters where a few thousand suffice at the alpine sites. An FNO reaches nonlocality in a single spectral layer and is therefore competitive on the marine case, but its spectral layers carry a parameter floor and produce no small-network configuration (Section~\ref{subsec:results_small}). The DahuNet, though it supplies the right local content, must still learn the propagation operator, which is iteration-hungry.

\subsection{Terminus caveat}
A secondary effect compounds this near the calving front. As grounded ice approaches flotation the surface flattens and~$\nabla s\to0$, so the local driving-stress proxy~$h^3|\nabla s|^3$ collapses toward zero. This is the leading-order shelf scaling behaving as expected---the freely-floating velocity vanishes with the local driving stress---but the true trunk speed does not, because it retains the offset and integration-length contributions that Appendix~\ref{app:feature_motivation} deliberately omits from the local factor: the velocity inherited at the grounding line and the stress transmitted over an upstream coupling length. The proxy therefore underestimates the flow most where the ice is fastest, and the local guess is partly pattern-inverted in the low-slope trunk. The operator the network must learn at Sermeq Kujalleq is thus propagation plus a near-terminus correction. This correction is larger than at the alpine sites, but remains smooth and low-complexity once the propagation is learned. 

Mitigating this directly is left for future work. Because the proxy is misleading only where the ice approaches flotation, a flotation-aware weighting of the driving-stress feature, or a treatment that handles grounded and floating domains separately, would suppress the anti-correlated signal at comparable cost; neither can supply the nonlocal grounding-line contribution, but such treatments become considerably more important for configurations with extensive floating ice, such as Antarctic ice-sheet simulations.

\subsection{Optimization budget versus representational capacity}

\begin{figure}
    \centering
    \includegraphics[width=0.375\linewidth]{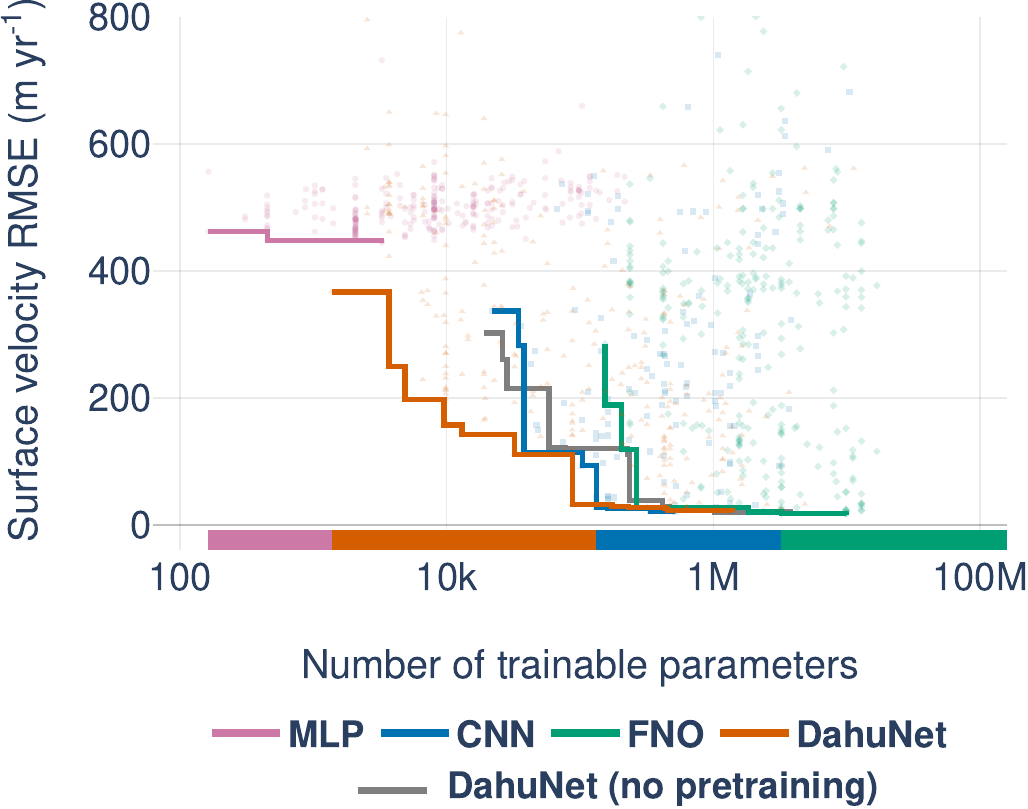}.
    \caption{Accuracy versus network size at Sermeq Kujalleq under a fixed pretraining budget. As in Fig.~\ref{fig:params}, thick staircase curves show the best surface-velocity RMSE attainable with at most a given number of trainable parameters; here every trial is additionally granted a fixed pretraining budget of~$5000$ optimizer iterations on the first stored state before the transient run, which is excluded from the reported parameter count. The gray staircase reproduces the DahuNet front without pretraining, from Fig.~\ref{fig:params}c, so that the two protocols can be compared at fixed architecture. Under the pretraining budget the DahuNet front extends down to a few thousand parameters and becomes the best option across~$10^4$--$10^5$ parameters, mirroring the alpine behavior of Fig.~\ref{fig:params}a,~b.}
    \label{fig:app_sermeq_pretrain}
\end{figure}

Two observations separate budget from capacity. First, capacity is not the binding constraint: in the transfer test, DahuNet solves the single marine state well from scratch (Table~\ref{tab:transfer_medians}), so the parametrization can represent the solution. Second, budget is. Figure~\ref{fig:app_sermeq_pretrain} repeats the accuracy-versus-parameters comparison of Fig.~\ref{fig:params} for Sermeq Kujalleq, granting every trial a fixed pretraining budget on the first stored state before the transient run. With this budget the DahuNet front extends down to a few thousand parameters and becomes the best option across~$10^4$--$10^5$ parameters, mirroring the alpine behavior. The obstacle in the no-warm-up protocol is therefore the optimization budget available to learn the propagation operator, not the network's capacity to represent the solution.

\section*{Declaration of competing interests}

There are no competing interests.

\section*{CRediT authorship contribution statement}

\textbf{\textsf{Thomas Gregov:}} Conceptualization, Methodology, Software, Writing – original draft, Writing – review and editing. \textbf{\textsf{Sebastian Rosier:}} Conceptualization, Methodology, Software, Writing – review and editing. \textbf{\textsf{Brandon Finley:}} Software, Writing – review and editing. \textbf{\textsf{Andreas Vieli:}} Funding acquisition, Supervision, Writing – review and editing. \textbf{\textsf{Guillaume Jouvet:}} Conceptualization, Funding acquisition, Methodology, Software, Supervision, Writing – review and editing.

\section*{Acknowledgements}

The authors acknowledge financial support from the Swiss National Science Foundation (SNSF) through the PINNACLE project (project number:~10002401). The authors would also like to thank Ludovic R\"{a}ss for his technical support in the use of GPUs at the Université de Lausanne.

\section*{Data availability}

The IGM software is available as an open-source code at \url{https://github.com/instructed-glacier-model/igm} and its documentation can be found at \url{https://igm-model.org/}.

\bibliographystyle{apalike}
\bibliography{ref}

\end{document}